\documentclass[preprint,prb]{revtex4-1}
\usepackage{graphicx,xcolor}
\usepackage{dcolumn}
\usepackage{bm}
\usepackage{amsfonts,amssymb,amsmath,geometry, url}
\usepackage{gensymb}
\usepackage{graphicx}
\usepackage{hyperref}
\usepackage{lineno}
\begin{document}
%

\title{On the origin of Blue Luminescence in Mg doped GaN} %
%
\author{Sanjay Nayak}
\affiliation{Chemistry and Physics of Materials Unit, Jawaharlal Nehru Centre for Advanced Scientific Research (JNCASR), Bangalore-560064, India}
\author{Mukul Gupta}%
\affiliation{UGC-DAE Consortium for Scientific Research, University Campus, Khandwa Road, Indore 452001, India}
 \author{Umesh V. Waghmare}
 \affiliation{Theoretical Sciences Unit, Jawaharlal Nehru Centre for Advanced Scientific Research (JNCASR), Bangalore-560064, India}
\author{S.M. Shivaprasad}
\email{smsprasad@jncasr.ac.in}
\affiliation{Chemistry and Physics of Materials Unit, Jawaharlal Nehru Centre for Advanced Scientific Research (JNCASR), Bangalore-560064, India}

\pacs{ }
\keywords{GaN:Mg, XANES, Defect complex, FEFF, SIESTA}
\date{\today}
\begin{abstract}
We uncover the origin of blue luminescence (BL) peak in Mg doped GaN thin film  using a combination of experimental  X-ray absorption near edge spectroscopy (XANES), \textit{first-principles} calculations based on density functional theory and full multiple scattering  theoretical analysis of various possible defect complexes and their XANES signatures. We demonstrate that a defect complex composed of  Mg substituted at Ga site (Mg$_\mathrm{Ga}$) and Mg at interstitial site (Mg$_\mathrm{i}$) is primarily responsible for  the observed BL by Donor-Acceptor Pair transition (DAP) associated with a deep donor state in the gap. It correlates with a higher (lower) oxidation state of N (Ga) in heavily Mg doped GaN than in its pristine structure, evident in our experiments as well as calculations. Physical and chemical mechanisms identified here point out a route to achieving efficient p-type GaN.  
\end{abstract}

\maketitle

\section{Introduction}
Gallium Nitride (GaN) is a wide band gap semiconductor (E$_\mathrm{g}$ = 3.51 eV at 2K) used for a wide variety of applications such as solid state lighting\cite{Dupuis2008}, high power and high frequency devices\cite{Sheppard1999}, and lasers\cite{Lu2008}. It is  essential  to have  both n and p-type semiconducting GaN to develop GaN/InGaN  quantum well (QW) based optoelectronic devices. Despite intensive research in the last few decades, some material issues related to GaN  remain open to be resolved. One of them is the unintentional n-doped behavior of native GaN; it is not clear whether it arises from point defects or from incorporation of impurities\cite{VanDeWalle2004a}. Although GaN based LEDs have already been  commercialized, their optimal performance is yet to be realized, partly because of the difficulties in efficient incorporation of p-type carriers in the host. Till date, magnesium (Mg) is the only p-type dopant to have been successfully employed in fabrication of p-GaN. However, a high activation energy (or ``ionization energy") of Mg in GaN \cite{Zhang2010} ($\approx$ 200 meV) requires  relatively high concentration of Mg in fabrication of efficient p-GaN. Secondly, higher Mg incorporation in GaN leads to the formation of point defects and/or defect complexes, and consequent self-compensation in Mg doped GaN\cite{Smorchkova2000}.  
\par
Incorporation of Mg in GaN results in characteristic luminescence peaks in Photo-luminescence (PL) and Cathodo-luminescence (CL) spectra depending on the dopant concentration. A small amount of Mg doping ($ < 10^{19}$ $cm^{-3}$) in GaN results in luminescence peaks  at 3.270 and 3.466 eV (at T=2K)\cite{Monemar2014}. At higher Mg dopant concentration ($ > 10^{19}$ $cm^{-3}$) in GaN thin films, a dominant PL peak appears in the range of 2.70 - 2.95 eV, emitting blue luminescence (BL), the origin of which has been  debated in the literature in the past decade\cite{Nonoda2016,Eckey1998,Reshchikov1999,Oh1998,Kaufmann2000,Hautakangas2006,Hautakangas2005}. Recent  DFT-based calculations  suggest that the emergence of BL is  due to  different ionization energies of nitrogen vacancies (V$_\mathrm{N}$) \cite{Buckeridge2015} and of hole localization at neighboring N atoms \cite{Lyons2012}. This however fails to explain the absence of BL from the samples with lower Mg-concentration. Moreover, no direct experimental evidence has been reported yet corroborating these mechanisms.
\par
X-ray absorption near edge spectroscopy (XANES) is an effective probe to determine the element or site specific properties such as oxidation state of an atom, local geometry and the electronic structure\cite{Wende2004}. Here, we report  growth of undoped and Mg doped GaN thin films and their characterization with  PL and XANES spectroscopies. We analyze the experimentally acquired XANES spectra by  co-rrelating them with  results of first-principles simulations of  various defect complexes and decipher the characteristic  features. Further, we use electronic structure of the relevant defect complexes obtained from DFT-based simulations to identify the luminescence centres  in Mg doped GaN.
\section{Methods}
\subsection{Experimental Details}
The films of  GaN in  Nanowall Network (NwN) geometry studied here were grown over (0001) plane of Sapphire by plasma assisted molecular beam epitaxy (PAMBE, SVTA-USA) system. The temperature of Gallium (Ga) effusion cell is maintained at 1030 $^o$C. A constant nitrogen flow rate of 8 sccm (standard cubic centimeter per minute), substrate temperature of 630 $^o$C, plasma forward power of 375 W and growth duration of 4 hours were used in growth of all the films. Mg and Ga fluxes are obtained from the beam equivalent pressure (BEP) and are varied by controlling  K-cells temperature. Other growth related details can be found elsewhere\cite{nayak2016gan}. We present analysis of  three samples of GaN films here: one pristine (A) and two doped (namely B and C), where the sample C is grown with higher Mg flux (Mg:Ga=0.1102)  than that used for B (Mg:Ga=0.0393). Optical properties of these films are studied with PL spectroscopy  at an excitation wavelength of 325 nm.  XANES spectra of these samples  are recorded in total electronic yield (TEY) mode  at SXAS beam line (BL-01) of the Indus-2 Synchrotron Source at Raja Ramanna Centre for Advanced Technology (RRCAT), Indore, India, in an ultrahigh vacuum (UHV) chamber with a base pressure of $10^{-10}$ Torr. Optical system of the beamline contains toroidal  mirror to focus the beam on sample surface (vertically as well as horizontally). The slit width before monochromator and sample are 1mm and 0.1 mm, respectively. The energy resolution in the acquired spectra is better than 0.2 eV. The details  of the experimental setup for XANES measurements are given in reference \onlinecite{phase2014development}. A typical data reduction procedure (background  removal  and  normalization)  of  the  XANES  spectra  is performed  using  the  Athena  software  package \cite{ravel2005athena}.

\subsection{Simulation Details}
In the simulation of XANES spectra, we have carried out first-principles DFT calculations using a combination of many codes. First, we  used the SIESTA code\cite{Soler2001} to obtain the  optimized atomic structure of defect configurations, where a Local Density Approximation (LDA) of Ceperley and Alder\cite{Ceperley1980} with Perdew and Zunger \cite{PhysRevB.23.5048} parametrization of the exchange and correlation energy functional was used.  Integrations over Brillouin Zone of $w$-GaN were sampled on a $\Gamma$- centered $5\times5\times3$ uniform mesh of k-points in a unit cell of reciprocal space\cite{Pack1977}. We relax positions of all  atoms  to minimize energy using a conjugate-gradients algorithm until the forces on each atom is less than 0.04 eV/\AA. Our optimized lattice parameters of the pristine GaN are `a' = 3.173 \AA \space and `c' = 5.163 \AA, \space which are in good agreement with the experiment\cite{leszczynski1996microstructure} (`a '= 3.186 \AA \space and `c' = 5.189 \AA). In simulation of GaN with defect(s), we  considered a $4\times4\times2$ supercell (128 atoms).  From the relaxed atomic structure obtained from SIESTA calculations we  constructed clusters in calculations to determine the \textit{ab-initio} XANES spectra  with FEFF9.05 code\cite{Rehr2010}, where the Hedin-Lundqvist exchange potential with an imaginary part of 0.2 eV is used and an exchange core hole is treated according to the final state rule in simulation of K-edges. We calculated  atomic potential for a 128 atoms cluster with a radius of 8 \AA\space and determined  the full multiple scattering XANES spectra by increasing the radius to 12.5 \AA \space around the absorber. To determine electronic structure and gap states of  the relevant defect configurations we used HSE06\cite{heyd2003hybrid} hybrid functional as implemented in VASP code \cite{Kresse1996}. In these calculations, we used  optimized lattice parameters and relaxed atomic structure obtained from SIESTA  calculations (details of the numerical parameters used in VASP calculations are given in the section I of the Supplementary Information).

\section{Experimental Results}
\begin{figure}[h]
   \centering
       \includegraphics[width=8cm]{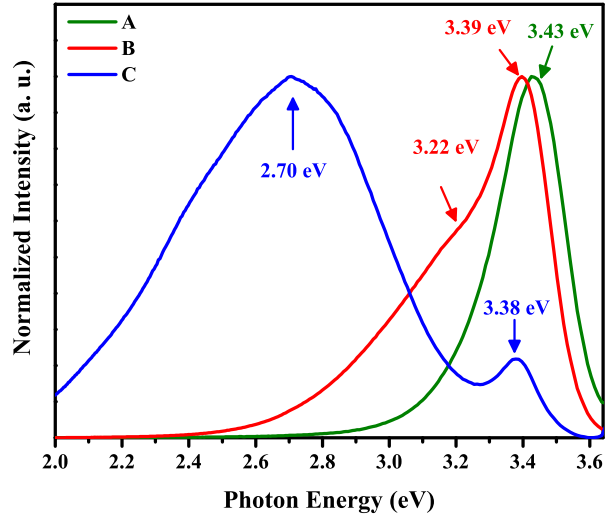}
    \caption{Normalized room temperature Photoluminescence (PL) spectra of  samples A, B and C.}
    \label{PL}
 \end{figure}

Details of the structural properties of the  films studied here were  discussed thoroughly in our earlier publication\cite{nayak2016gan}. As mentioned earlier, we have chosen GaN NwN as the host in the present work because of the superior optical properties of the film. It is well known that\cite{nayak2016gan,nayak2017} GaN in NwN geometry do not show any other luminescence peak(s) except near band edge (NBE)  ($\approx$ at 3.4 eV)  making it the preferred  choice here.  In the  photoluminescence spectra obtained at RT  (see Fig. \ref{PL}), it is seen that the sample A has only one dominant luminescence peak centered at 3.43 eV, assigned NBE emission of GaN, whereas the sample B exhibit two distinct peaks, centered at 3.39 and 3.22 eV, respectively. While the peak centered at 3.39 eV is assigned to NBE of  GaN, the peak centered at 3.22 eV is assigned to electron-Acceptor (e-A) or donor acceptor pair (DAP) transition in Mg-doped GaN\cite{Monemar2014}. PL spectra of sample C shows two distinct luminescence features, and they are identified as the NBE at 3.38 eV and the intense BL peak centered at 2.70 eV respectively. 

 \par
\begin{figure}[h]
   \centering
       \includegraphics[width=10cm]{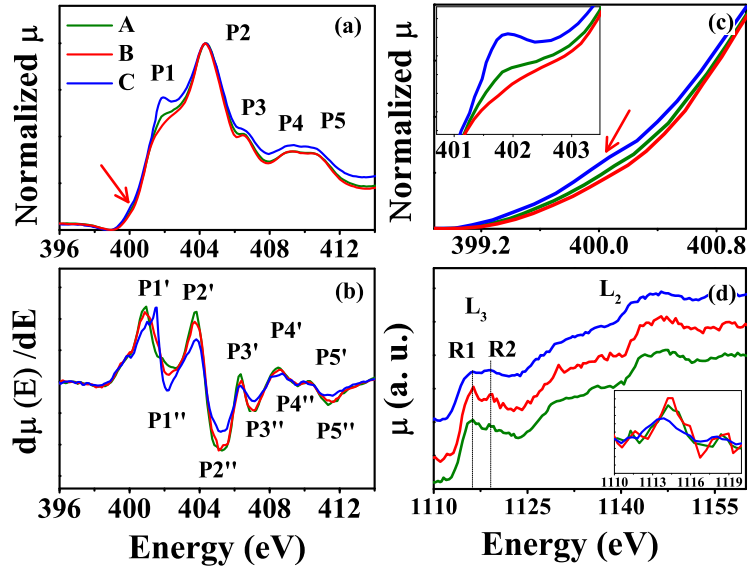}
    \caption{XANES spectra of normalized N K edges (a) and Ga L$_{2,3}$  edges (d) and their first derivatives at (b) and inset of (d), respectively. Figure (c) shows the enlarged version of the lower energy side of the absorption profile. Inset of Fig. (c) shows enlarged version of the feature P1.}
    \label{xanes}
 \end{figure}

To understand the origin of  different luminescence peak as observed in the PL spectra of samples, it is useful to study the electronic structure of all samples. To this end, we have probed the N K-edge and Ga L$_{2,3}$-edge of  the three samples using XANES (see Fig. \ref{xanes}(a) and (d)). For N K-edge, five distinct features (P1-P5) are seen clearly, which are consistent with earlier observations\cite{Lambrecht1997}. We have considered the absorption edge as the location of the first significant peak in the first derivative of  absorbance ($\mathrm{\mu (E)}$) with respect to energy (\textit{i.e.} $\mathrm{d\mu(E)/dE}$)\cite{Bittar2011} (see Fig. \ref{xanes}(b)). For each peak observed in Fig. \ref{xanes}(a) the corresponding maxima (P1' -P5') and minima (P1''-P5'')  are shown in Fig. \ref{xanes}(b). It is seen that the N K-edges (P1') of A and  B do not show any significant changes in their absorption threshold, whereas that of sample C shows a shift of $\approx$ 0.7eV towards higher energy relative to that of A,  suggesting  a small increase in the oxidation state of N atoms in sample C. Along with a change in the absorption edge, we  also observe a clear distortion of the peak P1. The peak intensity of P1 of sample C is higher and more pronounced than that of samples  A and B (see inset of Fig. \ref{xanes}(c)). This increase in the peak intensity of P1 with higher Mg incorporation is consistent with an earlier report\cite{Pan2001b}.  Increase in the intensity of the feature P1 may arise from the localized states that form due to higher Mg incorporation in the film. We further observe a small absorption feature at 400.16 eV (see arrow in Fig. \ref{xanes}(a) and  Fig. \ref{xanes}(c)).
Recorded L$_{2,3}$ edge spectra of Ga atoms are presented in Fig. \ref{xanes}(d), with the first derivative of L$_3$ edge in the inset.  Similar to N K-edge, we do not observe any changes in the absorption threshold of samples A and B. However, we  observe a small red shift ($\approx$ 0.80 eV) in that of sample C,  indicating  reduction in the oxidation states of Ga atoms in sample C. The two distinct features R1 and R2 are seen clearly in all three samples. For samples A and B, the intensity of R1 is higher than that of R2, whereas  for the sample C, intensity of R2 is higher than that of R1, with  a flat absorption profile near the L$_3$ edge. 
\par
\section{Theoretical Analysis and Discussion}
\subsection{X-ray Absorption Near Edge Structure}
\begin{figure}[h]
   \centering
       \includegraphics[width=10cm]{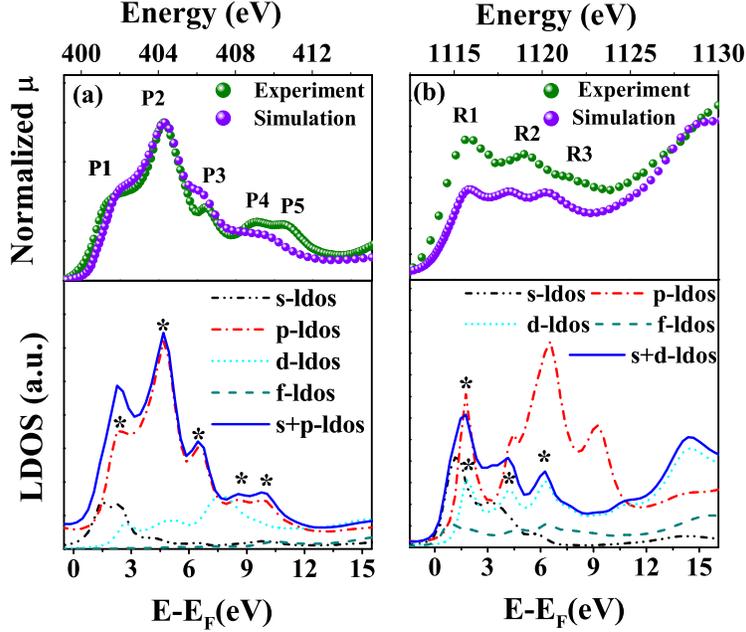}
    \caption{XANES spectra of N K-edge (a) and Ga L$_3$-edge (b) obtained from experiment (for commercial GaN) and simulation based on full potential multiple scattering methods. LDOS of individual orbital is shown at the bottom with their respective absorption edges.}
    \label{simulation}
 \end{figure}

To uncover the origin of  observed changes and features in experimental XANES spectra as a function of Mg doping concentration, we have obtained \textit{ab-initio} XANES spectra. We  focus on three  possible mechanisms, which are widely speculated to be  the origins of the BL in literature: (i) nitrogen vacancy complexes, (ii) the configuration with hole localization and (iii) Mg interstitial defect complex (ball and stick model of the different defect configurations are shown in Fig.S1 of section III of the Supplementary Information). Before studying the relevance of vacancy complexes to XANES spectra, we benchmarked the simulation parameters with careful analysis of the pristine GaN. Fig.\ref{simulation} (a) and (b) show theoretical  XANES spectra of N K-edge and Ga L$_3$ edge respectively, along with the experimentally observed spectra from a flat GaN epitaxial layer. Clearly, there is a good agreement between our theory and experiment (see Fig. \ref{simulation}), as well as the results reported earlier\cite{Moreno2007}.  
\par
We  observe that  N K-edge absorption edge is primarily dominated by the unoccupied N-2p orbitals, whereas  Ga L$_3$ edge has a strong `s+d' hybridized orbital character. Further, we  find that the feature R1 has a strong `s+d' hybridized character, and R2 and R3  are predominantly `d' character. The calculated local density of states (LDOS) projected on s, p, and d orbitals of N and Ga atoms reveal narrow energy sub-bands (marked with ``$\ast$" ) in agreement with the energies of the characteristic features seen in the XANES spectra.

\begin{figure}[h]
   \centering
       \includegraphics[width=10cm]{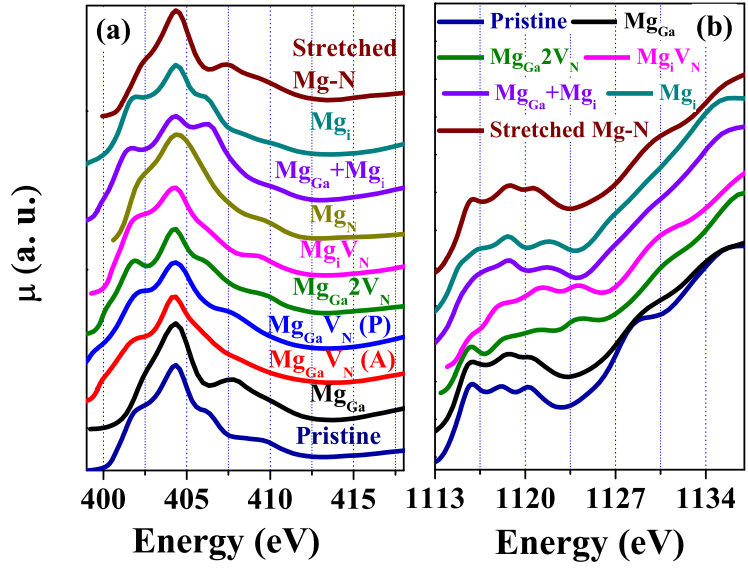}
    \caption{Simulated XANES spectra N K-edge (a) and Ga L$_3$- edge (b) for various defect configurations.}
    \label{defect}
 \end{figure}

 The simulated N K and Ga L$_3$  edges in XANES spectra of the  substitutional Mg at Ga site (Mg$_\mathrm{Ga}$)  are shown in Fig. \ref{defect}(a) and (b) respectively. We find that  peak P1 obtained from the simulation of  configuration with (Mg$_\mathrm{Ga}$) defect is not very prominent, and has a lower intensity relative to the pristine GaN. This reduction in  intensity of P1 is consistent with the behavior shown by  sample B, and we infer that no other defect complexes are present notably in sample B and the luminescence peak centered at 3.22 eV is due to recombination of electron-Acceptor pair (e-A).  Mulliken population analysis from results of SIESTA calculations suggests a small increase in the oxidation states of both 1$^{st}$ nearest neighbor (NN) N ($\approx$ 0.05 $|e|$) atoms as well as 1$^{st}$ NN Ga ($\approx$ 0.016$|e|$) atoms, that co-ordinate the site of Mg substituent in Mg-doped GaN relative to that of undoped  GaN (see Table \ref{tab}). Despite this small increase in  oxidation states of N and Ga atoms seen in our simulations, we do not observe a significant change in absorption edges of the sample B, due to significantly lower incorporation of Mg in the host. 
\par
Further, we simulated several defect configurations such as (i) complexes of Mg$_\mathrm{Ga}$ with a single N vacancy (Mg$_\mathrm{Ga}\mathrm{V_N}$), with two N vacancies (Mg$_\mathrm{Ga}-\mathrm{2V_N}$), with Mg  at interstitial site (Mg$_\mathrm{Ga}\mathrm{+Mg_i}$) (ii) complex of Mg  at interstitial site with a nitrogen vacancy (Mg$_\mathrm{i}-\mathrm{V_N}$), (iii) Mg at Nitrogen site (antisites) (Mg$_\mathrm{N}$),  and (iv) Mg at interstitial site (Mg$_\mathrm{i}$). The characteristic signatures of these configurations in the XANES spectra are shown in Fig. \ref{defect}(a) and (b). In the configurations of Mg$_\mathrm{Ga}\mathrm{V_N}$, we have considered N-vacancies in the axial and basal planes of the $w$-GaN, as  the four Ga-N bond lengths are not same for GaN$_\mathrm{4}$ tetrahedra. We find that  former configuration is energetically more preferable, and do not observe any significant change in the characteristics of XANES spectra with respect to pristine GaN. Thus, the increase in the peak P1 of sample C cannot be attributed to  formation of a Mg$_\mathrm{Ga}\mathrm{V_N}$ complex.
 	
 \par
 
 A comparison between the experimental and simulated results for XANES spectra  clearly suggests that the possible causes for the enhanced intensity of  P1feature in N K-edge of the sample C can be Mg$_\mathrm{Ga}-\mathrm{2V_N}$, Mg$_\mathrm{i}-\mathrm{V_N}$, Mg$_\mathrm{Ga}\mathrm{+Mg_i}$ or $\mathrm{Mg_i}$. Mulliken charges of  N atoms, near these defect complexes due to formation of defect complexes are listed in Table \ref{tab} along with the formation energies. Clearly the oxidation states of N (Ga) atoms increase (decrease) slightly in  these defect configurations. Thus, a clear conclusion on determination of dominant defect(s) could not be reached from the Mulliken population analysis alone. Further, our estimates of the formation energies of  these defect complexes obtained  using Zhang-Northrup scheme \cite{Zhang1991} (see Table \ref{tab} and details on the method used for calculation  of defect formation energy in section II of Supplementary Information) reveal that the defect configuration Mg$_\mathrm{Ga}\mathrm{+Mg_i}$ has the lowest formation energy, and  is probably the most preferable defect complex that should form in Mg-doped GaN. Recently, Miceli \textit{et al.}\cite{miceli2016self} and Reshchikov  \textit{et al.}\cite{reshchikov2014green} predicted from Hybrid functional based DFT calculations that Mg interstitial is the energetically  preferable defect in Mg-doped GaN, which was neglected earlier due to over-estimation of its formation energy with a semilocal functional\cite{van1999defects,dupuis1996gallium}, which is consistently
evident in the results of our calculations here (see Tab.\ref{tab}).  Simulated L$_3$ edge spectra of the configurations Mg$_\mathrm{Ga}\mathrm{+Mg_i}$ and $\mathrm{Mg_i}$  show (see Fig. \ref{defect}(b)) that the peak R2 has a higher intensity than R1, as observed experimentally only for the sample C. 
\par
  Thus, we propose that the observed increase in  intensity of P1 of sample C is due to the  increase in the unoccupied donor states, originating from the formation of  defect complex Mg$_\mathrm{Ga}\mathrm{+Mg_i}$ and/or $\mathrm{Mg_i}$. Further the shift in the absorption threshold of N-K  edges of sample C in comparison to sample A is due to  reduction in the Mulliken charges (oxidation states) of N atoms of the MgN$_4$ tetrahedra. We find the relaxed atomic structure of  GaN containing the Mg$_\mathrm{Ga}\mathrm{+Mg_i}$ defect complex shows  elongation of axial  Mg-N bond by 14\%, while in basal plane  one Mg-N bond in the basal plane contracts  by 2.5\%  while other two Mg-N bond stretches by 5.4\% relative  to the  Ga-N bonds of pristine GaN.
 \par 
To connect with the prediction of Van de Walle \textit{et al.},\cite{Lyons2012} we   simulate the XANES spectra of N K and Ga L$_3$ edges  by stretching the Mg-N bond to a value 15\% higher than the Ga-N bond, while allowing other atoms to relax (see Fig. \ref{defect}). We do not see any significant change in the N K-edge w.r.t. pristine GaN. Thus, the distortion of peak P1 in Fig.\ref{xanes} (a) can not be attributed to longer Mg-N bond. A careful observation of the simulated N K-edge spectra in Fig. \ref{defect}(a) reveals an absorption edge (at $\approx$ 400 eV) for all the N-vacancy related complexes thus  we attribute this feature at $\approx$ 400.16 eV in the XANES spectra (see Fig.\ref{xanes}(c)) to  unoccupied states associated with N-vacancies.

\begin{table*}
\centering
\caption{Neutral Defect Formation Energy of different defect complexes at N-rich condition. The Mulliken charges of $1^{st}$ nearest neighbor (NN) of the defect complex with reference to pristine GaN. The -ve (+ve) sign in Mulliken charge indicates increase (decrease) in the oxidation state. }
\label{tab}
\begin{tabular}{c|c| cc}
\hline
\hline \
    {Defect complex}         & {Formation Energy}         & \multicolumn{2}{c}{Mulliken charge in $|e|$}   \\
    \cline{3-4}
        
        { }     & (in eV)         & \space NN N      & NN Ga     \\

          \cline{3-4}
          \hline 
          { $\mathrm{Mg_{Ga}}$}     & {1.20}         & {-0.050}      & { -0.016 }     \\
          {$\mathrm{Mg_{Ga}V_N}$ }     & {0.51}         & {-0.015}      & {+0.152}     \\
          { $\mathrm{Mg_{Ga}2V_N}$}     & {4.91}         & {-0.046}      & {+0.181}     \\
          {$\mathrm{Mg_{i}V_N}$}     & {8.47}         & {-0.056}      & {+0.242}     \\
          {$\mathrm{Mg_{N}}$ }     & {9.78}         & {-0.064}      & {+0.074}     \\
          {$\mathrm{Mg_{Ga}+Mg_i}$ }     & {-0.17}         & {-0.061}      & {+0.055}     \\
          {$\mathrm{Mg_i}$ }     & {4.80}         & {-0.015}      & {+0.068}     \\

\hline
\hline
\end{tabular}
\end{table*}
\subsection{Electronic Energy Levels}
As DFT-LDA typically underestimates the band gap (band gap of GaN calculated with SIESTA is 2.06 eV, much lower than its experimental value of 3.51 eV (see section IV of the Supplementary Information)), we used hybrid HSE06 functional based calculations to determine energy levels of the defect states in the electronic structure using VASP code (see Fig.\ref{band}). In these simulations with a $4\times4\times2$ supercell, we used only $\Gamma$-point in sampling the Brillouin Zone integrations. Our  estimate of the band gap of pristine GaN is 3.36 eV, reasonably close to the experimentally observed band gap of 3.51 eV at T=2K and 3.43 eV  observed at RT in this study. For the configuration Mg$_\mathrm{Ga}$, we find a shallow acceptor state ($\approx$ 0.22 eV  above the VBM), which has a predominant  N-2p orbital character. Configurations with (Mg$_\mathrm{Ga}\mathrm{+Mg_i}$) and Mg$_\mathrm{i}$ exhibit deep donor states in the electronic gap $\approx$ at 3.14 and 3.07 eV above VBM. Thus,  the transitions from the deep donor states to the shallow acceptor state occur at ($3.14 - 0.22 = 2.92$ eV), very close to the emission peak of BL  (here 2.7 eV) observed in sample C (see Fig.\ref{band}). We note that the concentration of Mg in our simulations of Mg-doped GaN is higher than experiment, and this small difference is partly due to that.
\par
 \begin{figure}[!htb]
   \centering
       \includegraphics[width=10cm]{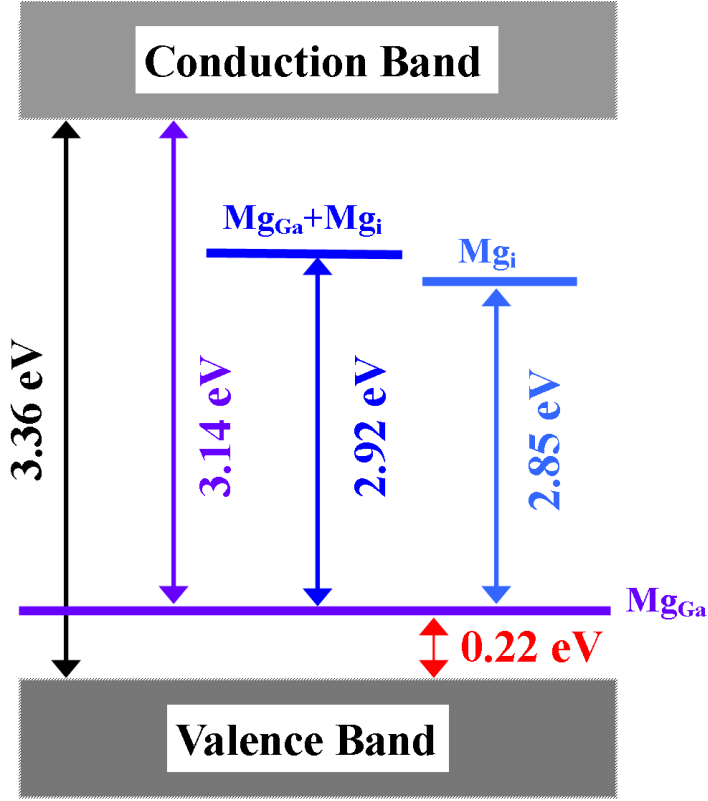}
    \caption{A schematic of electronic structure of heavily Mg doped GaN obtained with hybrid  HSE06 functional based calculations (at $\Gamma$ point).}
    \label{band}
 \end{figure}
 Although in past, a similar  mechanism (transition between deep donor to shallow acceptor) on the origin of BL has been proposed, its atomistic origin has not been clear\cite{Teisseyre2000,Reshchikov1999} . Lyons \textit{et al.}\cite{Lyons2012}  claimed an alternative mechanism where BL is a result of  transitions of electrons from the conduction band to the deep and  localized Mg$_{\mathrm{Ga}}$ acceptor level. However, it fails to explain large shifts in the BL peak with increase in excitation intensity and absence of thse same in lightly doped samples. In addition, work of Buckeridge \textit{et al.} \cite{Buckeridge2015} suggests that the BL may be due to the formation of isolated N-vacancy where authors used a hybrid quantum mechanical (QM) and molecular mechanical (MM) embedded cluster method. In contradiction to this, reports claimed that  isolated N vacancy gives rise to yellow luminescence (2.18 eV) \cite{yan2012role} and/or green luminescence (2.35 eV) in Mg doped GaN \cite{reshchikov2014green}. Also, some  issues related to the accuracy in  calculations of  Buckeridge \textit{et al.}\cite{Buckeridge2015} have been reported by other groups\cite{PhysRevLett.115.029701,lyons2015first}. Recently, Wahl \textit{et al.}\cite{Wahl2017} studied the site occupancy  of Mg in GaN by implanting radioactive Mg in GaN, and found a notable amount of Mg in interstitial sites,  while the majority of them occupy the substitutional Ga sites.  As mentioned earlier, some reports \cite{miceli2016self,reshchikov2014green} suggest the formation energy of Mg occupying at interstitial site is less in p-type GaN. Thus,  Mg may prefer to be at the interstitial sites in the films during the epitaxial growth process. 
\par 
As the epitaxial growth temperature of GaN is reasonably high (630 $^o$C in the present work), mobility of Mg adatom is high during growth. When Mg  at interstitial site diffuses close to a  Ga vacancy site in the process, it takes up the same and becomes substitutional Mg at Ga site in GaN\cite{Wahl2017}. In the films grown under the low Mg flux, occurrence of Mg$_\mathrm{Ga}$ is expected to be dominant, resulting in  formation of a shallow acceptor state in the electronic gap, and associated 3.22 eV  peak in the luminescence spectra. In contrast, during the epitaxial growth with higher Mg flux,  the number of available Ga vacancy sites are not abundant enough for diffusing  Mg$_\mathrm{i}$ to get converted to Mg$_\mathrm{Ga}$. Instead, Mg$_\mathrm{i}$ adatoms will  pair up with other suitable defects due to their relatively low formation energy and form defect complexes such as Mg$_\mathrm{Ga}\mathrm{+Mg_i}$, which create a deep donor state in the electronic gap. Our work shows that these are responsible for the BL emission arising from the deep donor to shallow acceptor state transitions. Further, the mechanism proposed in this study  on the origin of BL clearly explain the the large shift of peak position with increase in excitation intensity and absence of BL in lightly doped samples. Based on our analysis, we  propose that the synthesis of p-type GaN under lower Ga flux (Ga poor condition) will be an efficient way relative to  higher flux.
\section{Summary}
We have uncovered  the origin of observed BL in Mg doped GaN through a combination of experiments and theoretical analysis of Mg incorporated at different concentration GaN thin films. With clear evidence in PL and XANES of heavily Mg incorporated  GaN films, we show that the observed BL originates from defect complexes formed of  interstitial Mg (Mg$_\mathrm{i}$) and substitutional Mg (Mg$_\mathrm{{Ga}}$). The BL is associated with a transition from the deep donor state in electronic gap to shallow acceptor state.  Our experiments reveal a slightly higher oxidation state of N and lower oxidation state of Ga in heavily Mg incorporated GaN than those in  pristine GaN, which are supported well by our first-principles calculations.
\section*{Acknowledgments}
The authors thank Professor C. N. R. Rao for his support and guidance. SN acknowledges DST for a Senior Research Fellowship.  The authors gratefully acknowledge  JNCASR, UGC-DAE CSR and RRCAT, Indore for providing facilities. UVW acknowledges support from a JC Bose National Fellowship and a TUE-CMS project funded by Nano Mission, Department of  Science and Technology, Government of India.
 \bibliographystyle{apsrev4-1}
\bibliography{xanes} 
\clearpage
 \section*{Additional information}
 \subsection{Numerical parameters used in VASP calculation}
First-principles DFT calculations  were carried out using a plane-wave projector augmented wave (PAW)  method as implemented in the VASP code, where a Local Density Approximation (LDA) of Ceperley and Alder is used for the exchange and correlation energy functional. The reference valence elecronic configurations of Ga, N and Mg were considered as $3d^{10}4s^24p^1$, $2s^22p^3$ and $3s^23p^0$, respectively. We used an energy cutoff of 500 eV to truncate plane wave basis.  Electronic energy spectrum at $\Gamma$ point was calculated by using  Heyd-Scuseria- Ernzerhof (HSE) hybrid functional, where the mixing parameter for the Hartree-Fock exchange potential is set at 25\%. The screening parameter in HSE calculations is fixed at 0.2.
\subsection{Procedure to calculate defect formation energy}
Formation energy of defects (for neutral state) in the bulk  was calculated using Zhang-Northrup scheme,  given by \\

 $E_f = E_{tot} (defect) - E_{tot} (pristine) + \Sigma n_i\mu_{i}$ \\
 
 where $ E_{tot} (defect)$ and $E_{tot} (pristine)$ are the total energies of super-cells containing a defect and the reference pristine  structure, respectively. $n_i$ and $\mu_i$ represent the number of atom added or removed (if atom(s) are added it will take positive sign where as if atom(s) are removed it will take -ve sign) and chemical potential of  $ i^{th}$ species, respectively. In this work, we have calculated the defect formation energy under  N rich conditions. Under N rich conditions $\mu_N$ is the energy of N- atom (obtained from the total energy $E_{tot}(N_2)$  of $N_2$ molecule, \textit{i.e.} $\mu _ N = \dfrac{1}{2} E_{tot}{ (N_2)} $).   The chemical potential of Ga is  calculated using the assumption of thermodynamic equilibrium, $\it {i.e.}$ $\mu_{Ga}+\mu_N = E_{GaN}[bulk]$; where $E_{GaN}[bulk]$ is the total energy of one formula unit of bulk \textit{w}-GaN. We have used chemical potential of Mg ($\mu_{Mg}$)  as the energy of single Mg atom in the hcp phase  ($\mu _ {Mg} = \dfrac{1}{2} E_{tot}{ (Mg _{hcp})} $, noting that the primitive unit cell of hcp structure contains two atoms.
 \clearpage
\subsection{Atomic Models Used for DFT and FMS calculation}
 \begin{figure}[!h]
   \centering
       \includegraphics[width=17cm]{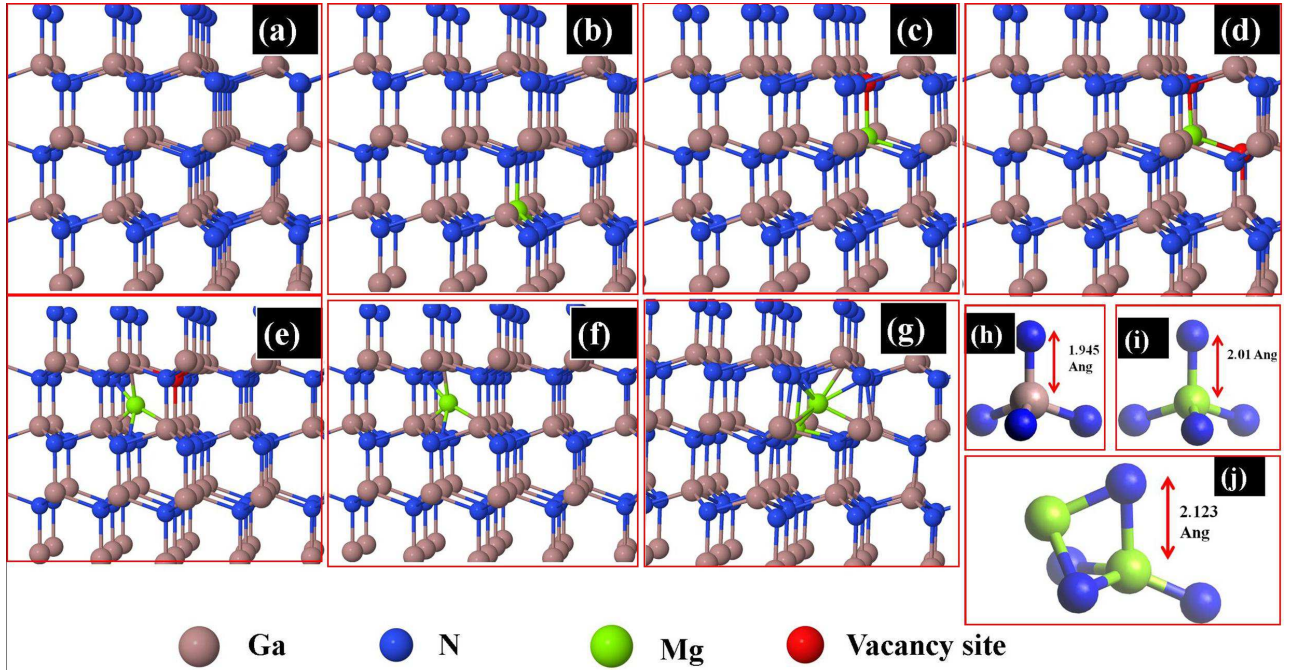}
    \caption{Ball and stick model of  different configurations used in DFT and multiple scattering theory calculations as Pristine w-GaN (a), $Mg_{Ga}$ (b), $Mg_{Ga}-V_N$ (c), $Mg_{Ga}-2V_N$ (d), $Mg_{i}V_N$ (e), $Mg_{i}$ (f), $Mg_{Ga}+Mg_i$ (g). Figure (h) and (i) shows tetrahedral structure of pristine Ga-N and Mg-N in Mg doped GaN, respectively. Figure (j) shows the relaxed structure of the $Mg_{Ga}+Mg_i$ defect complex.}
    \label{model}
 \end{figure}

\subsection{Electronic structure of  defect complexes}
 \begin{figure}[!h]
   \centering
       \includegraphics[width=16cm]{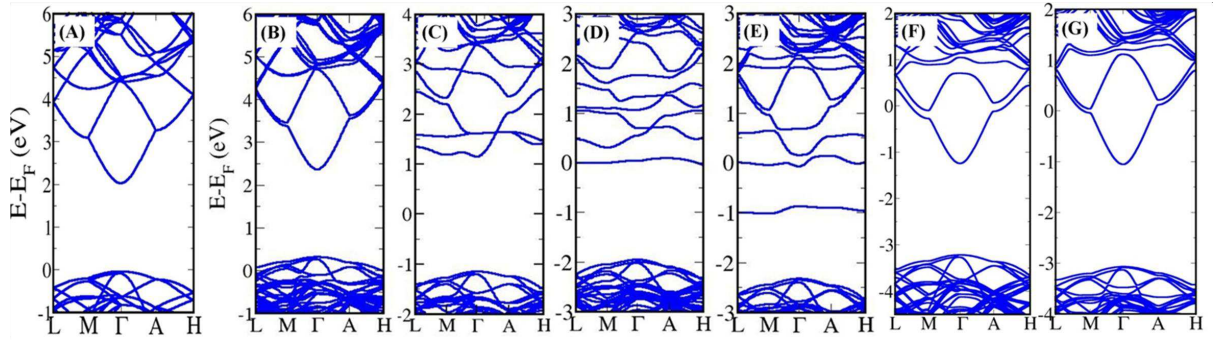}
    \caption{Electronic structure of different defect configurations obtained from SIESTA. Pristine w-GaN (A), $Mg_{Ga}$ (B), $Mg_{Ga}-V_N$ (C), $Mg_{Ga}-2V_N$ (D), $Mg_{i}V_N$ (E), $Mg_{i}$ (F), $Mg_{Ga}+Mg_i$ (G). The Fermi level is set at 0 eV. }
    \label{band}
    \end{figure}

\end{document}